# Self-cleaning effect in an all-fiber spatiotemporal mode-locked laser based on graded-index multimode fiber

Chuansheng Dai[1], Zhipeng Dong[1], Jiaqiang Lin[1], Peijun Yao[1,*], Lixin Xu[1], Chun Gu[1]

1. University of Science and Technology of China, Hefei 230026, China

*yap@ustc.edu.cn

Abstract: We demonstrate an all-fiber spatiotemporal mode-locked laser based on graded-index multimode fiber. Due to the high damage threshold of graded-index multimode fiber devices, the output single-pulse energy reach up to 11.67 nJ. The beam quality at mode-locked state is significantly improved compared with continuous wave state as a result of the self-cleaning effect. By amplifying the pulse in the multimode fiber amplifier, high power pulses with high beam quality can be obtained in the all-fiber system based on graded-index MMFs.

Keywords: Spatiotemporal mode-locked laser; Self-cleaning effect; All-fiber structure; Graded-index multimode fiber; High power ultrafast fiber laser.

## 1. Introduction

High power ultrafast fiber laser not only has the advantages of compact configuration, high efficiency and reliability, but also has high output power, ultrashort pulse width and high pulse energy. Therefore, it has application value in laser micromachining, optical communication, national defense, military and scientific research. However, the mode field area of single-mode fiber (SMF) limits the improvement of output power and pulse energy. The large mode area (LMA) fiber can effectively suppress the nonlinear effect in the fiber and improve the damage threshold of fiber devices which draw much attention of researchers. Most of the LMA fibers are multimode fibers (MMF) in practical applications, including step-index MMF, graded-index MMF and multi-core fibers. Recently, nonlinear propagation in graded-index MMF has attracted major interest [1-5]. Graded-index MMFs can support very rich modal interaction due to small modal dispersion which is comparable to chromatic dispersion. Nonlinear effects in MMF have been studied extensively such as multimode soliton formation [4, 6-7], beam self-cleaning effect [8-11], spatiotemporal instabilities [12-13], soliton self-frequency shift [14], nonlinear multimode interactions [15], multimode dispersive waves [16].

In the last few years, the first spatiotemporal mode-locking (STML) MMF laser [17] has been demonstrated by Wright et al. Multiple transverse and longitudinal modes have been locked by strong spatial and spectral filtering to achieved ultrashort pulses, which provides a new platform to research nonlinear wave propagation. It is well known that the beam quality of MMFs is very poor, the input laser beams of high spatial quality fade into irregular granularities called speckles, which to some extent limited its application. Fortunately, the nonlinear beam self-cleaning phenomenon for femtosecond pulses has been observed in graded-index MMF [8]. With the increase of pulse energy, the near-field beam profile changes from speckle pattern to centered, bell-shaped transverse structure. Recently, researchers found that space beam self-cleaning can achieve high efficiency with a few-mode excitation in graded-index MMFs [18]. Beam self-cleaning effect has enabled significant applications such as beam combining [9] and the development of high-power fiber laser.

In this letter, we bulid an all-fiber STML fiber laser based on graded-index MMFs. STML is realized by using nonlinear polarization rotation (NPR) technique, and spatial filtering is generated by direct fusion of fibers with different core sizes to compensating the modal- and chromatic-dispersion of MMFs. Stable mode-locking that operate at 1075.45 nm is achieved with a spectral width of 7.24 nm at 3 dB. The repetition rate is 17.29 MHz with an estimated pulse duration about 33.28 ps. The single-pulse energy reach up to 11.67 nJ. The beam quality at mode-locked state is significantly improved compared with continuous wave state. After further amplification by the MMF amplifier, the nonlinear beam self-cleaning effect is more significant, and we finally get high-power pulses with high beam quality. Our research has potential value in the field of high power ultrafast fiber laser.

## 2. Experimental setup

The experimental schematic of STML fiber laser is depicted in Fig. 1(a). A 3.2-meter-long ytterbium-doped fiber (YDF) (Liekki YB1200-10/125DC) which supports about three transverse modes is pumped by a 980nm multimode laser diode (LD) though a mutipump combiner. The pump stripper at the end of YDF is used to peel off the residual pump light. The pigtails of the combiner and the pump stripper are passive multimode fibers (Liekki 10/125DC) with 10 μm core. Other passive fibers in the ring cavity are graded-index MMFs (Corning OM4 50/125) with 50μm core which support ∼240 modes. The MMF shifts about 10μm from the center of the pigtails fiber of the pump stripper during fiber splicing in order to excite multiple transverse modes. In reverse, the large difference in the core size of the two fibers leads to strong spatial filtering when the signal light returns to the combiner, which can compensating the modal- and chromatic-dispersion of fibers. Two polarization controllers (PC) and a polarization-dependent isolator (PD-ISO) realize nonlinear polarization rotation (NPR) mode-locking. The output laser is emitted from the 10% port of the 90:10 optical coupler (OC1).

In order to further confirm the STML characteristics of the pulse, we set up a spatial sampling device as shown in Fig. 1(b). The 10% of the 90/10 coupler (OC2) was connected to a fiber collimator. By adjusting the relative position of two fiber collimators, different transverse mode components of the output beam can be selected for measurement. The other port of OC2 was connected to a multimode fiber amplifier to analyze the effect of pulse power on the self-cleaning effect as shown in Fig.1 (c).

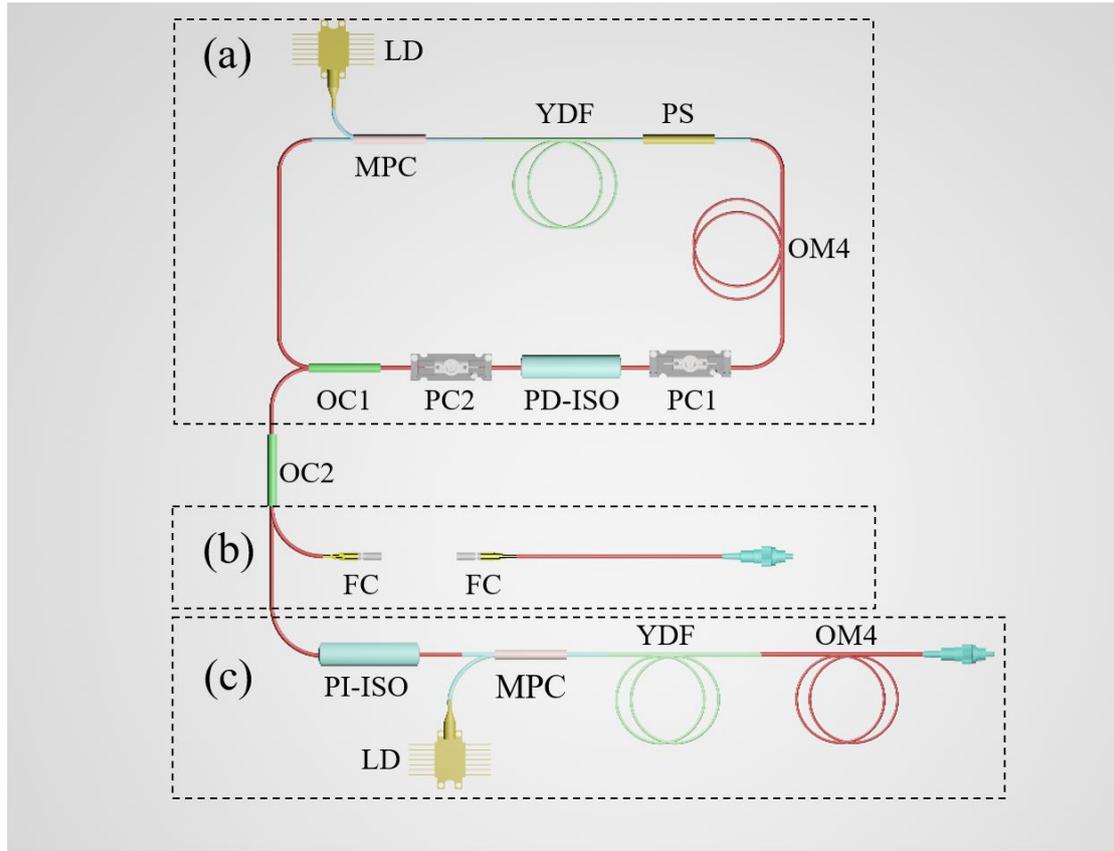

Fig. 1. Experimental setup. (a) All-fiber STML fiber laser (MPC, multipump combiner; PS, pump stripper; OM4, passive graded-index MMFs; PC, polarization controller; PD-ISO, polarization-dependent isolator; OC, optical coupler); (b) Spatial sampling device (FC, fiber collimator); (c) Multimode fiber amplifier (PI-ISO, polarization-independent isolator).

### 3. Results and discussions

When the pump power increases to 4.9 W, we can obtain STML pulses by adjusting the PCs. We have measured the pulse characteristics with 8.4 W pump power as shown in Fig. 2. The spectra is measured by an optical spectrum analyzer (ANDO AQ6317B) and showed in Fig. 2(a), where curve (e) shows the spectrum of the whole output. The STML operates at a center wavelength of 1075.45 nm with a width of 7.24 nm at 3dB. The curve (f)-(g) are the spectra of the beams sampled by the spatial sampling device as shown in Fig. 1(b). By adjusting the relative position of the two collimators, we can get the beams with different transverse modes components. Obviously, the spectra of the sampled beams are very different, since different transverse modes have different resonant frequencies [17]. Fig. 2(b) shows the frequency-domain characteristics which are measured by a 2 GHz radio-frequency (RF) spectrum analyzer (AV4021). Fig. 2(c) shows the time-domain characteristics which are measured by a 4 GHz oscilloscope (Teledyne LeCroy WaveRunner 640Zi) with a 3 GHz bandwidth photodetector. The corresponding RF spectra of (e)-(g) and pulse trains of (e)-(f) are shown in Figs. 2(b)-2(c). The repetition rate of the output pulse is 17.29 MHz with a signal to noise ratio of 46 dB, and the interval of output pulses is 57.84 ns. The RF spectra and pulse trains of the beams with different transverse modes components are almost the same, which confirm the STML characteristics [19]. The autocorrelation trace of the single pulse detected by an autocorrelator (APE PulseCheck SM250) is shown in Fig. 2(d). The fitting curve shows a 33.28 ps pulse duration with a Gaussian shape assumption. The time-bandwidth product (TBP) is 62.53, which is much larger than the minimum TBP of Gaussian pulse of 0.44. Therefore, there is a large chirp in the pulse.

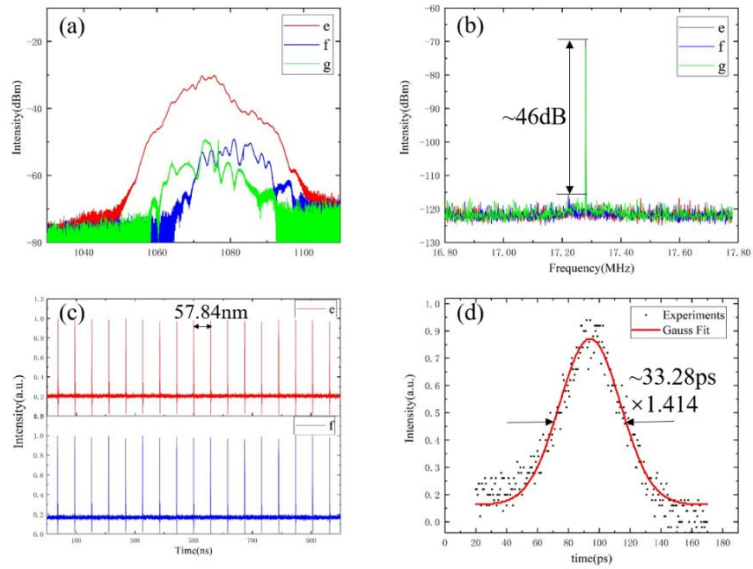

Fig. 2. Single-pulse mode-locking state: (a) and (b) Optical spectra and RF spectra of the whole output (curve e) and the spatial sampling outputs (curve f-g), respectively; (c) pulse trains of the whole output (curve e) and the spatial sampling output (curve f); (d) Autocorrelation trace of output pulses.

We have measured the output characteristics of the STML lasers at different pump powers, and the results are summarized in Fig. 3. The pump power range is from 4.9W to 12.8W, limited by the damage threshold of optical fiber devices. The output average power increased from 48.5 mW to 201.7 mW, while the single-pulse energy increased from 2.81 nJ to 11.67 nJ in the pump power range. As far as we know, this is the highest single-pulse energy obtained from an all-fiber STML laser. However, the slope effect is only about 1.9%. We think that the loss caused by spatial filtering is the main reason for low slope efficiency.

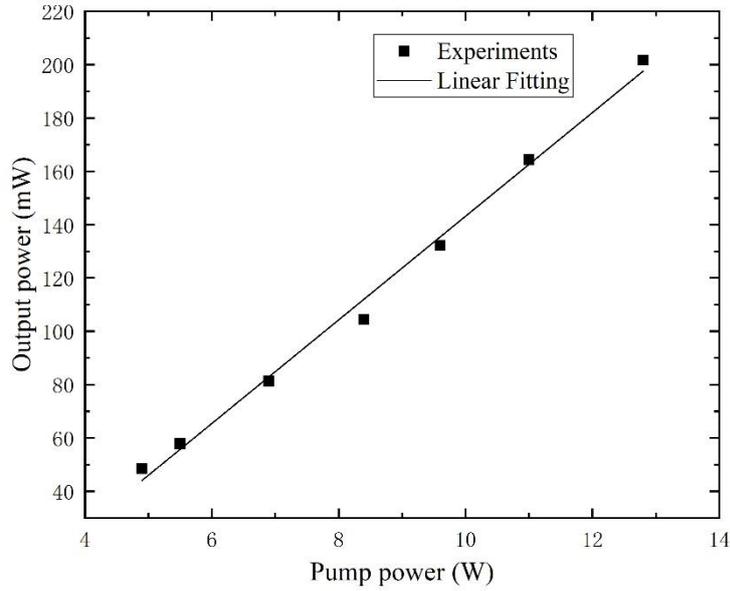

Fig. 3. Output average power as a function of the pump power.

For an input pump power of 8.4 W, a pulse output of 104.5 mW is obtained. The corresponding single-pulse energy and peak power are 6.04 nJ and 0.18 kW, respectively. Fig. 4 shows the profiles of the output beams which are imaged by using a CCD camera. By adjusting PCs, the measured beam profiles for continuous wave and STML operation case were demonstrated in Fig. 4(a) and Fig. 4(b), respectively. For the case of continuous wave operation, there were a lot of speckle in CCD image, which is the typical characteristic of a MMF beam profile [20]. When the laser operation was converted from continuous wave operation to STML operation, the output beam profile was significantly improved. This self-cleaning phenomenon is due to the intermodal interactions mediated by Kerr nonlinearity. Due to the limitation of damage threshold of optical fiber devices, we further amplify the STML pulse outside the cavity. The structure of the multimode fiber amplifier is shown in Fig. 1(c). A 5 m long ytterbium-doped fiber (YDF) (Liekki YB1200-10/125DC) is used as the gain medium of the amplifier. The amplified laser passed through a 20 m long OM4 fiber, which provide sufficient nonlinear interaction length. We used the output pulse of STML fiber laser as the seed pulse to incident into the amplifier. By changing the pump power of the amplifier, we can get the high power pulse with different average power. And the relationship between output average power and pump power is shown in the Fig. 5. The Fig. 4(c) and Fig. 4(d) shows the beam profiles from the amplifier with the output average power of 1 W and 3.5 W, respectively. Obviously, the beam profile of the amplified beam possesses a well-defined bell-shaped structure in the core center. With the increase of the average power, the size of the beam profile becomes smaller, which is due to the self-focus effect.

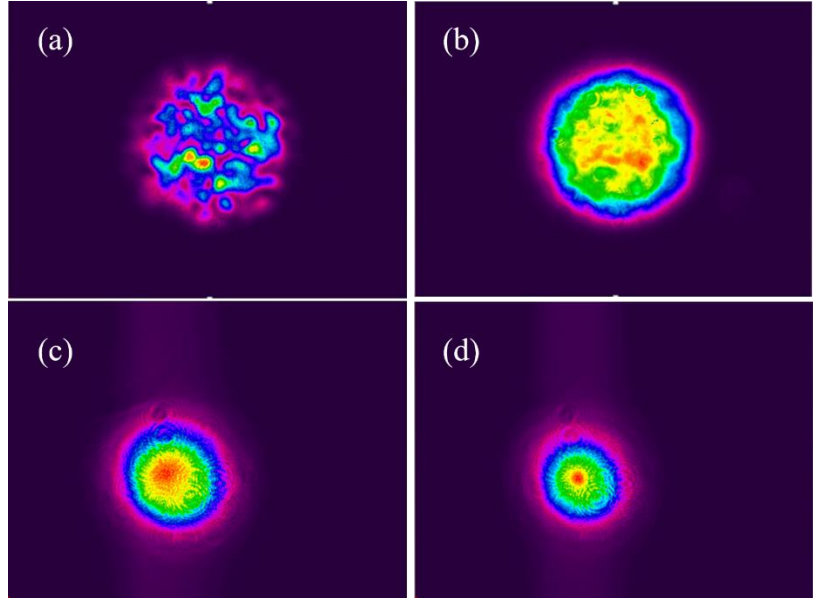

Fig. 4. Measured beam profiles from the all-fiber laser resonator for (a) continuous wave operation case and (b) STML operation case; measured beam profiles from the multimode fiber amplifier with the output average power of (c) 1 W and (d) 3.5 W.

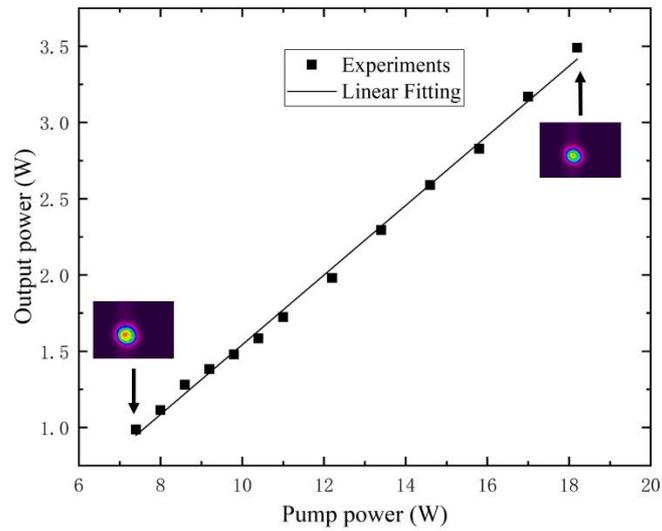

Fig. 5. Output average power of the amplifier as a function of the pump power (inset: measured beam profiles from the amplifier for different output average power).

## 4. Conclusion

In this letter we built an all-fiber STML laser based on graded-index MMFs. This oscillator generated ultrafast pulse with 33.28 ps pulse duration, 17.29 MHz repetition rate, and 46 dB signal to noise ratio, which operate at 1075.45 nm with a spectral width of 7.24 nm at 3 dB. Due to the high damage threshold of OM4 fiber devices, the average power and single-pulse energy can reach up to 201.7 mW and 11.67 nJ, respectively, which is the highest average power and single-pulse energy obtained from an all-fiber STML laser to our knowledge in the literature. The beam quality at mode-locked state is significantly improved compared with continuous wave state. By amplifying the pulse in the multimode fiber amplifier, high power pulses with high beam quality can be obtained in the all-fiber system based on graded-index MMFs. The all-fiber cavity design has the advantages of high stability, compact structure and low cost. We believe the proposed cavity has potential value in the field of high power ultrafast fiber laser for the high damage threshold, high pulse energy and high beam quality.

**Acknowledgements**

This work was supported by.We also thank the USTC Center for Micro and Nanoscale Research and Fabrication for supporting the experiment.